\newcommand{\Figdir}{.}

\documentstyle[aps,prl,multicol,rotate,epsfig,fancyheadings,amsbsy,amssymb]{revtex}

\begin{document}
\title{Photoemission Spectroscopy from Inhomogeneous Models of Cuprates}
\author{J. Eroles$^{1,2}$, G. Ortiz$^{1}$, A. V. Balatsky$^{1}$ and A.
R. Bishop$^{1}$}
\address{$^1$Theoretical Division, 
Los Alamos National Laboratory, Los Alamos, NM 87545, USA.\\
$^2$Centro At\'{o}mico Bariloche and Instituto Balseiro, 
S. C. de Bariloche, Argentina.\\}

\date{Received \today }

\maketitle

\begin{abstract}
We investigate the electronic dynamics in the underdoped cuprates focusing on the effects
of one-dimensional charge stripes. We address recent experimental Angular-Resolved 
Photoemission Spectra results on (La$_{1.28}$Nd$_{0.6}$Sr$_{0.12}$)CuO$_4$. We find that
various inhomogeneous models can account for the distribution of quasiparticle weights  
close to momentum  ${\bf k}=(\pi,0)$ and symmetry related points. The observed
flat dispersion region around the same ${\bf k}$ point can only be addressed by certain 
classes of those inhomogeneous models which locally break spin symmetry.
Homogeneous models including hopping elements up to second neighbors cannot reproduce the
experimental quasiparticle weight,
since most of it is centered around ${\bf k}=(\frac {\pi}{2},\frac {\pi} {2})$.
\end{abstract}


\vspace*{-0.4cm}
\begin{multicols}{2}

\columnseprule 0pt


Recent interest in the inhomogeneous states of high-$T_c$ superconductors has
gained further impetus from new experiments that support 
charged stripe formation. One set of data is available on the Nd-doped 
LSCO\cite{tranquada} and, more recently, a second set concerns
oxygen doped LSCO\cite{birgeno-1D}. Angular-Resolved
Photoemission Spectra (ARPES) obtained on the commensurate-doped Nd LSCO\cite{expe-shen}
revealed that the electronic spectra, believed to contain
static stripes, are consistent with the electronic states having 
quasi-one-dimensional (1D) character. 


The question that naturally arises from these measurements is: what
are the most appropriate models that can capture both the formation
of the stripes and adequately describe the spectroscopic features of
the electronic states in these compounds? The issue of the
physics that drives stripe formation is a matter of extensive current
debate\cite{formation-debate}, with a number of models that 
can describe the static and dynamical stripe formation. 

Faced with the new experimental data, we
choose here a different approach. We assume from the
beginning that the Hamiltonian for our model is inhomogeneous to mimic the 
effects of stripes. As a
consequence the electronic states will be quite different from the
conventional homogeneous models, such as a $tt'$-$J$ model. 
We find that suitable explicitly inhomogeneous models allow
a good fit to the ARPES data, and capture the main qualitative features
that are consistent with a quasi-1D nature of electronic states on the
stripes. 

The inhomogeneous models we will consider have 
already been introduced in Ref. \cite{ours}, where the
basic microscopic scenario starts from a homogeneous $t$-$J$
Hamiltonian as a reference background model:
\begin{equation} 
\label{H} 
H_{t\!-\!J} = -t  \sum_{\langle {\bf r,\bar{r}} \rangle, \sigma}
c^{\dagger}_{{\bf r} \sigma} c^{\;}_{{\bf \bar{r}} \sigma} 
+ J
\sum_{\langle {\bf r,\bar{r}} \rangle} ({\bf S}_{\bf r} \cdot {\bf
S}_{{\bf \bar{r}}} - \frac{1}{4} \bar{n}_{\bf r} \bar{n}_{{\bf
\bar{r}}} ) . 
\end{equation} 
We have also considered the $tt'$-$J$ model which includes hopping elements up to
second neighbors.
To mimic the stripe segments, we add specifically inhomogeneous {\it magnetic}
interactions. These inhomogeneous terms locally break translational invariance
and spin-rotational $SU(2)$ symmetry: 
\begin{equation} 
H_{\rm inh} = \sum_{\langle \alpha,\beta \rangle} \delta J_z \
S^z_\alpha S^z_\beta + \frac{\delta J_{\perp}}{2} \left( S^+_\alpha 
S^-_\beta + S^-_\alpha S^+_\beta \right).
\label{H-inh} 
\end{equation} 
Here $\delta J_{\perp} \neq \delta J_z$, represents the magnetic 
perturbation of a static, local, {\it Ising} anisotropy, locally lowering
spin symmetry (termed a $t$-$JJ_z$ model\cite{ours}). Only a few links (where the
stripes are located) have this lowered spin symmetry. The Ising
anisotropy is also sufficient to produce a spin gap\cite{ours} and pair binding of
holes in this class of models is substantial \cite{ours}. An
alternative scenario for stripe segments can be realized through
inhomogeneous terms of the form 
\begin{equation} 
\tilde{H}_{\rm inh} = -\sum_{i} \epsilon_i (1 -
\bar{n}_{i}) \ ,
\label{H-inh2} 
\end{equation} 
where $\epsilon_i > 0$ represents an on-site orbital energy which is
non-vanishing only at the stripe sites $i$ ($tt'$-$J\epsilon_i$ model). 
ARPES signal of the noninteracting electron model
with external potential, consistent with the stripe pattern, has been previously addressed in
Ref.~\cite{salkola}.
The spin rotation
symmetry is not broken in this case. This kind of model resembles, although is different, 
 to that studied in Ref.\cite{tohyama}, where it was shown to
reproduce experimental neutron magnetic
scattering, optical conductivity and the ARPES 
measured broad spectrum  near  ${\bf k} =
(\frac{\pi}{2},\frac{\pi}{2})$ along the direction $(0,0)-(\pi,\pi)$\cite{tohyama}.

In the rest of this paper the energy
scale will be determined by $t$. As we will vary  the magnitude of the spin 
anisotropy\cite{ours}, we choose to fix $J/t=1$. This is
 a little higher than the commonly assumed value 
for homogeneous cuprate systems of
$J/t=0.4$, but close enough to be physical. 
It is not our purpose here to define an energy scale which can be
related to the experiments, but rather to demonstrate that the main aspects of the
experiments can be address by particular kinds of models. 
Systematic tuning of the parameters
will be published elsewhere.

\begin{figure}[tbp]
\epsfxsize=7cm
$$
\centerline{\epsfbox{\Figdir/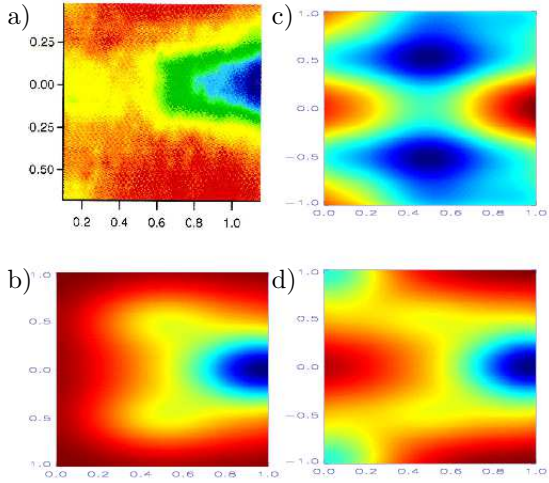}
\put(-207,175){a)}
\put(-207,75){b)}
\put(-107,175){c)}
\put(-107,75){d)}
}
$$
\caption{(a) Experimental ARPES from {\protect \cite{expe-shen}}.
(b), (c) and (d) 
correspond to the unsymmetrized calculated spectrum for  $t$-$JJ_z$, 
$tt'$-$J$ and $tt'$-$J\epsilon_i$ models, respectively. Note that no integral of spectral weight has been
taken yet.}
\label{non-symmetrized} 
\end{figure}

\begin{figure}[tbp]
\epsfxsize=8cm
$$
\centerline{\epsfbox{\Figdir/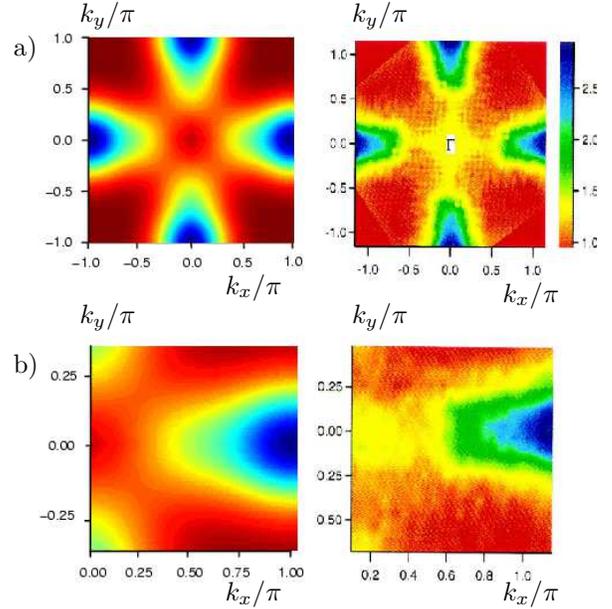}
\put(-220,210){a)}
\put(-220,90){b)}
\put(-195,223){$k_y/\pi$}
\put(-140,120){$k_x/\pi$}
\put(-195,109){$k_y/\pi$}
\put(-140,4){$k_x/\pi$}
\put(-35,117){$k_x/\pi$}
\put(-92,223){$k_y/\pi$}
\put(-35,4){$k_x/\pi$}
\put(-92,109){$k_y/\pi$}
}
$$
\caption{Comparison between the inhomogeneous $t$-$J$$J_{z}$ model 
($t/J=1$ and $\delta J_{\perp}=-0.9$) integrated up to $\Delta \omega = 0.13 t$ (left)
and ARPES
experimental results (right) from {\protect \cite{expe-shen}} integrated over $\Delta
\omega=100$meV. 
(a) Four
quadrants view (b). Zoom of the zone around $(\pi,0)$.}
\label{Comp-exp-100} 
\end{figure}

\begin{figure}[tbp]
\epsfxsize=8cm
$$
\centerline{\epsfbox{\Figdir/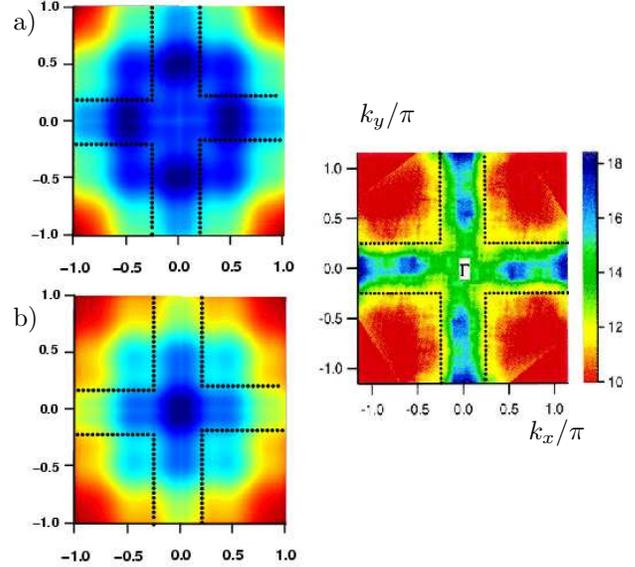}
\put(-230,205){a)}
\put(-230,93){b)}
\put(-35,50){$k_x/\pi$}
\put(-99,170){$k_y/\pi$}
}
$$
\caption{Comparison between the inhomogeneous $t$-$J$$J_{z}$ model 
($t/J=1$ and $\delta J_{\perp}=-0.9$) (left) and ARPES
experimental results (right){\protect \cite{expe-shen}}
integrated in a 500 meV range. 
Four quadrants view. (a)$\Delta \omega = 2.5t$; (b)$\Delta \omega = 3.5t$.}
\label{Comp-exp-500} 
\end{figure}

Recent ARPES experiments in
La$_{1.28}$Nd$_{0.6}$Sr$_{0.12}$CuO$_4$ seem to indicate a
1D electronic structure \cite{expe-shen}. 
At low transferred energies and integrating the spectral weight within a 100 meV
window ($\Delta \omega=$100 meV), the
quasiparticle weight looks like that depicted  
in Figs.~\ref{non-symmetrized} and \ref{Comp-exp-100}. 
Basically, there is almost no weight at ${\bf k} =
(\frac{\pi}{2},\frac{\pi}{2})$, and 
 most of the weight is concentrated near the point ${\bf k} = (\pi,0)$. Moreover, one 
can see from
the quasiparticle band structure that there is almost no
dispersion in $k_y$. Besides, there is a flat band for large $k_x$ that
can be related to a velocity scale\cite{sasha}. 
By integrating the same data over a window 
of 500 meV ($\Delta \omega=$500 meV), the experimental spectrum is like that in 
Fig.~\ref{Comp-exp-500}. There is no peak at ${\bf k} =
(\frac{\pi}{2},\frac{\pi}{2})$ and most of the spectral weight is transferred from 
${\bf k} = (\pi,0)$ to
${\bf k} = (\frac{\pi}{2},0)$ and symmetrically equivalent points. Both cases ($\Delta
\omega=100$ meV or 500 meV) are characteristic of a 1D behavior of the
electronic degrees of freedom\cite{expe-shen}.
These experiments were all performed at a commensurate hole filling corresponding to $x=1/8$. 

\begin{figure}[tbp]
\epsfysize=6cm
$$
\centerline{\rotate[r]{\epsfbox{\Figdir/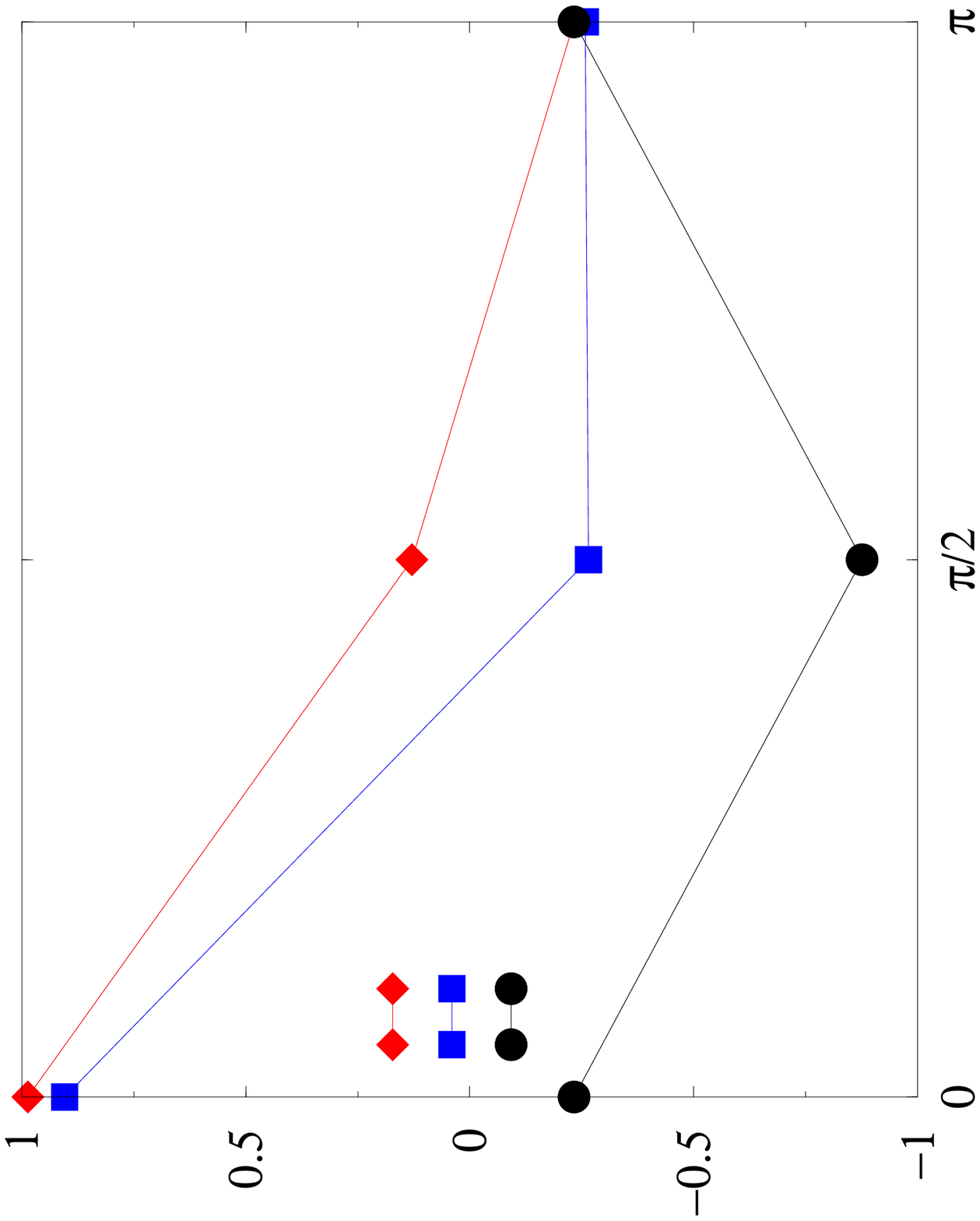}}
\put(-133,80){{\scriptsize{ $tt'$-$J$}}}
\put(-133,72){{\scriptsize{ $t$-$JJ_z$ }}}
\put(-133,64){{\scriptsize{ $tt'$-$J\epsilon_i$ }}}
\put(-110,80){\epsfbox{\Figdir/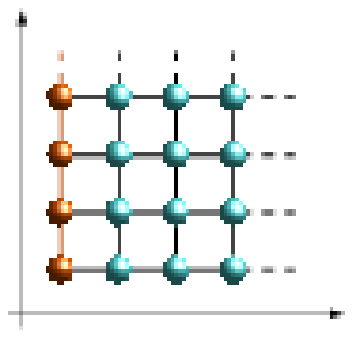}}
}
$$
\caption{ Dispersion relation for the different models along the  $(0,0)-(\frac {\pi}
{2},0)-(\pi,0)$ line. The experimentally observed flat band close to 
${\bf k}=(\pi,0)${\protect \cite{expe-shen}} is well reproduced by the $t$-$JJ_z$ and
$tt'$-$J$ models.}   
\label{fig-disp}
\end{figure}

Our goal here is to study different homogeneous and inhomogeneous 
models and compare to the
experimental dynamic properties measured in Ref.\cite{expe-shen}. We used 
the Lanczos numerical exact diagonalization method on a 4$\times$4 cluster (see inset Fig.~\ref{fig-disp}).
When stripes
are present, the Hamiltonian along one leg is modified by adding the inhomogeneous terms (like in
Eqs.~\ref{H-inh} or \ref{H-inh2}).

The spectral function we need to compute is given by

\begin{equation}
{\cal A}({\bf k},\omega) = \sum_n | \langle \Psi_n^{\nu +1} |
c^{\dagger}_{{\bf k} \sigma} | \Psi_0^{\nu} \rangle |^2 \ \delta(\omega
-(E_n^{\nu +1} - E_0^{\nu})) \ ,
\end{equation}
where $| \Psi_n^{\nu} \rangle$ corresponds to an eigenstate of the
Hamiltonian in the subspace of $\nu$ holes with energy $E_n^{\nu}$. The first
poles of the corresponding Green's function define the quasiparticle weight through the intensity,
and the energy associated to that quasiparticle through the position of the
poles. As the experiments in Ref.~\cite{expe-shen} are  at $x=1/8$, we
carried out the simulations with one hole added to a background where there was
already one hole present ($\nu=1)$. For our 16 sites cluster, the final concentration
then corresponds to $x=1/8$.

Since our system is finite, we can only consider a finite set
of ${\bf k}$ values in the Brillouin zone. These are: ${\bf k} =
(k_x,k_y)$ with $k_{x,y} = 0, \pm \frac{\pi}{2}, \pm \pi$. To compare
with experimental results we associate a Gaussian with a small width ($\sigma
\simeq 0.1$) centered at each computed ${\bf k}$ point and with a height 
corresponding to the
spectral weight for that  ${\bf k}$ point.
Finally, we sum all these
Gaussians centered at the ${\bf k}$ values of our finite
lattice. The resulting field is color coded with blue corresponding to maxima and
red to minima. For each graph the maximum is scaled to 1.

In Fig.~\ref{non-symmetrized} we show (a) the experimental spectrum from Ref.~\cite{expe-shen}
and the corresponding unsymmetrized model predictions
 for (b) $t$-$JJ_z$, (c) $tt'$-$J$ and (d) $tt'$-$J \epsilon_i$ models.
For the calculations we have taken the first quasiparticle pole for each ${\bf
k}$ point. (Note that for some ${\bf k}$ points the weight is very small).
Clearly predictions from the 
inhomogeneous models ((b) and (c)) are qualitatively different from the homogeneous
$tt'$-$J$ model. In both inhomogneous cases the weight is concentrated around ${\bf k}=(\pi,0)$ as in the
experimental ARPES.  
But if stripes are present, the experiment averages domains with different
stripe orientations.  Therefore, in order to compare with the experiments, we
symmetrize the numerical results by taking the
average with the rotated spectrum and integrating the results in a frequency window $\Delta
\omega$ for each ${\bf k}$ point, as was done in Ref.~\cite{expe-shen}.


\begin{figure}[tbp]
\epsfverbosetrue
\epsfxsize=7cm
\epsfysize=9cm
$$
\centerline{\epsfbox{\Figdir/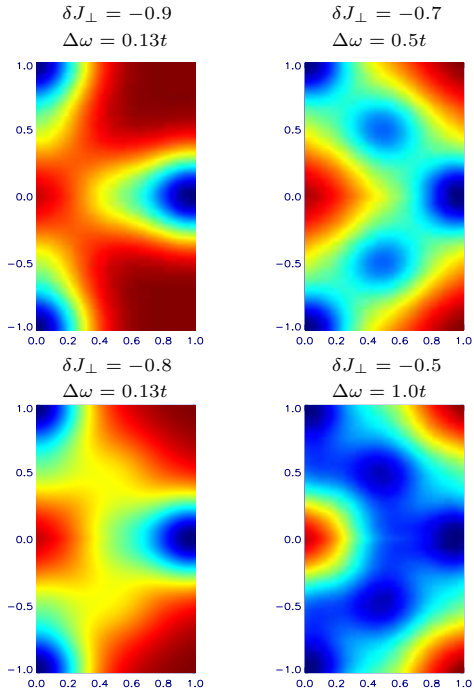}
\put(-172,260){{\scriptsize{ $\delta J_{\perp}=-0.9$ }}}
\put(-172,250){{\scriptsize{ $\Delta \omega=0.13t$ }}}
\put(-172,127){{\scriptsize{ $\delta J_{\perp}=-0.8$ }}}
\put(-172,117){{\scriptsize{ $\Delta \omega=0.13t$ }}}
\put(-70,260){{\scriptsize{ $\delta J_{\perp}=-0.7$ }}}
\put(-70,250){{\scriptsize{ $\Delta \omega=0.5t$ }}}
\put(-70,127){{\scriptsize{ $\delta J_{\perp}=-0.5$ }}}
\put(-70,117){{\scriptsize{ $\Delta \omega=1.0t$ }}}
}
$$
\caption{  Spectra for different parameters and integration windows for the $t$-$JJ_z$
model. 
As the strength of the stripe is decreased ($\delta J_{\perp} \rightarrow 0$, 
making the model more homogeneous) noticeable weight is transfered to  
${\bf k}=(\frac {\pi} {2},\frac {\pi} {2})$ and to the boundary of the ``ghost''
Anti-Ferromagnetic Brillouin zone. }
\label{fig-Jz}
\end{figure}

In Fig.~\ref{Comp-exp-100} we compare the calculated spectra in the inhomogeneous
$t$-$JJ_z$ model (Eq.~\ref{H-inh}), where the stripes are represented as local {\it Ising}
anisotropy, symmetrized and integrated over a range up to $\Delta \omega =$100 meV. The
Hamiltonian  corresponds to Eqs.~\ref{H} and \ref{H-inh} with $\delta J_{\perp}=-0.9$. 
Figure \ref{Comp-exp-100} corresponds to (a) the full 4 quadrants view, and (b) to a
partial view around ${\bf k} = (\pi,0)$. The agreement is very good.   In the experiment
significant  spectral weight was found around ${\bf k} = (0,0)$. Our data,  on the other
hand, do not show any significant weight at ${\bf k} = (0,0)$. It is well known that the
reduction to one band effective models is not very good at this ${\bf k}$ point\cite{Z}.
Thus the lack of weight at ${\bf k}=(0,0)$ is a consequence of the
reduction to the $t$-$J$ model and not important for our present considerations.

In
Fig.~\ref{fig-Jz} we display the quasiparticle weights for various values of the stripe
anisotropy. Note the transfer of weight from ${\bf k} = (\pi,0)$ to ${\bf k} = (\frac
{\pi} {2},\frac {\pi} {2})$ as the system becomes more homogeneous. 


Quasiparticle features involving many sites (and therefore with higher
energies) will not be properly captured in our small cluster. Nevertheless, it still
seems able to describe the main trends as the window of
integration in the experiment is raised to  $\Delta \omega =$ 500 meV. As seen
in Fig.~\ref{Comp-exp-500}(a) the weight concentrates around ${\bf k} = (\frac
{\pi} {2},0)$ and equivalent points, even though the weight at  ${\bf k} = (\frac
{\pi} {2},\frac {\pi} {2})$  is still significant. As the window of
integration is raised to $\Delta \omega =$ 3.5$t$, most of the weight is
concentrated between ${\bf k} = (0,0)$ and  ${\bf k} = (\frac {\pi} {2},0)$.
Probably because of the size of our cluster,  none of the models studied can 
exactly reproduce 
the features of this high energy profile; indeed as expected, they are all very
similar in this range.

\begin{figure}[tbp]
\epsfxsize=8cm
$$
\centerline{\epsfbox{\Figdir/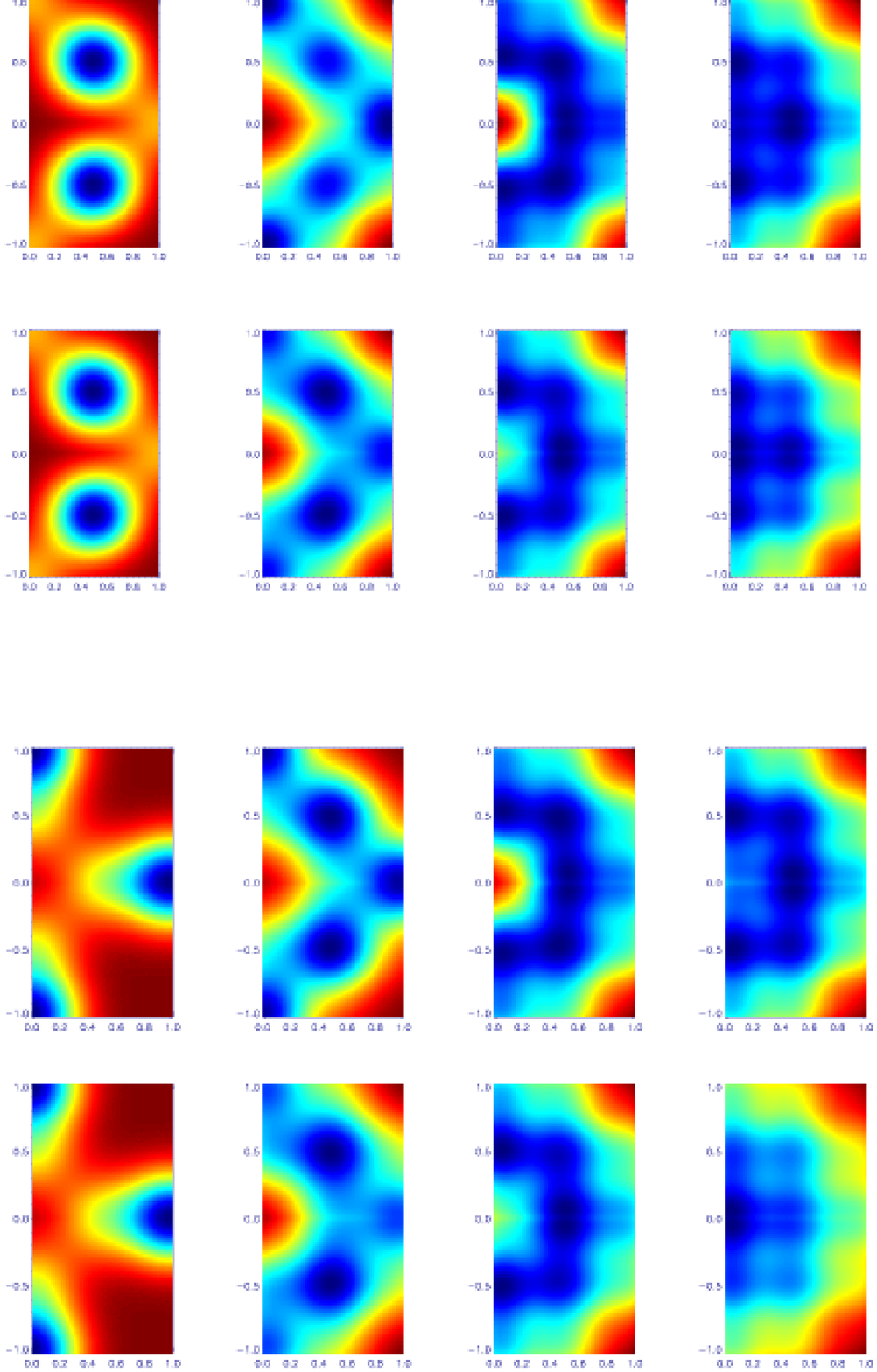}
\put(-220,350){{\scriptsize{ $\Delta \omega = 0.01$ }}}
\put(-220,268){{\scriptsize{ $\Delta \omega = 0.1$ }}}
\put(-161,350){{\scriptsize{ $\Delta \omega = 0.5$ }}}
\put(-161,268){{\scriptsize{ $\Delta \omega = 1.0$ }}}
\put(-102,350){{\scriptsize{ $\Delta \omega = 1.5$ }}}
\put(-102,268){{\scriptsize{ $\Delta \omega = 2.0$ }}}
\put(-44,350){{\scriptsize{ $\Delta \omega = 2.5$ }}}
\put(-44,268){{\scriptsize{ $\Delta \omega = 3.0$ }}}
\put(-230,180){-----------------------------------------------------------------------}
\put(-235,150){{\normalsize{ b) }}}
\put(-235,340){{\normalsize{ a)}}}
\put(-220,163){{\scriptsize{ $\Delta \omega = 0.01$ }}}
\put(-220,79){{\scriptsize{ $\Delta \omega = 0.1$ }}}
\put(-161,163){{\scriptsize{ $\Delta \omega = 0.5$ }}}
\put(-161,79){{\scriptsize{ $\Delta \omega = 1.0$ }}}
\put(-102,163){{\scriptsize{ $\Delta \omega = 1.5$ }}}
\put(-102,79){{\scriptsize{ $\Delta \omega = 2.0$ }}}
\put(-44,163){{\scriptsize{ $\Delta \omega = 2.5$ }}}
\put(-44,79){{\scriptsize{ $\Delta \omega = 3.0$ }}}
}
$$
\caption{ Comparison of the spectral density between (a) homogeneous
$tt'$-$J$ ($t'=-0.3t$) and (b) inhomogeneous $t$-$J$$J_{z}$ 
($t/J=1$ and $\delta
J_{\perp}=-0.9$) models. Blue corresponds to maximum intensity and red to
minimum. Each plot corresponds to the indicated range of integration 
$\Delta \omega$ (in units of $t$). }
\label{Comp-Jz-tJ}
\end{figure}

In Fig.~\ref{Comp-Jz-tJ} we compare the distribution of quasiparticle weights  for 
homogeneous $tt'$-$J$ and inhomogeneous $t$-$JJ_z$ models after symmetrization and for
different integration windows $\Delta \omega$. We   study the  $tt'$-$J$ model since it is
well known that, in our cluster, the bare $t$-$J$ has a hidden symmetry for 1 hole that makes ${\bf
k}=(\pi,0)$ and ${\bf k}=(\frac{\pi}{2},\frac{\pi}{2})$ degenerate. 
In particular for a chain, the
dispersion relation for a $t'$ term is $\epsilon_{k}=-2t' \cos(2k)$, and therefore it
lowers the quasiparticle energy at both $k=0$ and, more importantly, at $k=\pi$. If for
some reason the system has a 1D (or quasi 1D) behavior, this $t'$ term will transfer
quasiparticle weight from  $k=\frac {\pi} {2}$ to $k=\pi$, as observed in the experiments.
From Fig.~\ref{Comp-Jz-tJ} it can be seen that  in the
homogeneous model the experimental weights cannot be
reproduced by $t'/t=-0.3$ (the commonly assumed value for $t'$\cite{armando}) 
for any $\Delta \omega$. 
It is also interesting to note that in the
inhomogeneous $tt'$-$J\epsilon_i$  model, after symmetrization, 
the prominent peak at 
${\bf k}=(\pi,0)$ and equivalent points (as in Fig.~\ref{Comp-Jz-tJ}(b) for 
$\Delta \omega=0.01t$ and $\Delta \omega=0.1t$)  is 
comparable to ${\bf k}=(\frac {\pi} {2},\frac {\pi} {2})$ for most of the 
$\Delta \omega$ window sizes. Therefore the agreement with experiments is poor. 
For large $\Delta \omega$ all the models are very much alike. 
This is to be expected since if the stripe perturbation is small enough, only the low
energy states change significantly, and therefore all inhomogeneous and
homogeneous models give the same qualitatively spectra for sufficiently large
$\Delta \omega$.

%
%
%



As  mentioned earlier, another distinctive feature in the ARPES experiments is the
flat dispersion around ${\bf k}=(\pi,0)$. This flat region may be associated with a
velocity which in turn defines the linear relation between the separation between stripes
and the superconducting transition temperature\cite{sasha}. 
Therefore it is important to see if the flat region is present in our
models. In Fig.~\ref{fig-disp} we show the position of the first pole for each model, for
the ${\bf k}$ points along the $(0,0)-(\frac {\pi} {2},0)-(\pi,0)$ line. The  $t$-$JJ_z$ 
and $tt'$-$J$
models are flat enough to be consistent with the experiment, but as we have seen above, 
the latter
cannot account for the weight distribution (see Fig.~\ref{Comp-Jz-tJ}). 
On the other hand the model labeled as
$tt'$-$J\epsilon_i$ does not show the flat dispersion. This is a further indication that
even though inhomogeneities are necessary to account for the ARPES experiments, not all
kinds of inhomogeneities are appropiate.

In conclusion, we have calculated the ARPES response in several inhomogeneous and homogeneous
models. We used exact diagonalization in a cluster with $C_{4v}$ symmetry and hole doping 
appropiate to $x=\frac {1} {8}$. We found that the experimentally measured quasiparticle peak
near ${\bf k}=(\pi,0)$ and symmetry equivalent points is well reproduced only by the
inhomogeneous models. We also find that for all models, some weight peaked at ${\bf k}=(\frac
{\pi} {2},\frac {\pi} {2})$.   It is of comparable intensity (although usually smaller) to the
peak at ${\bf k}=(\pi,0)$, and is present for  some integration windows. Of those models we
studied, only those that locally break $SU(2)$  symmetry can also explain the experimentally
observed  flat dispersion around this same ${\bf k}$ point. We note that this same  model has
a spin-gap, substantial binding of holes and exhibits an incommensuration in the spin
structure factor\cite{ours}. It thus appears to capture many important experimental  features
in the electronic, magnetic and potentially superconducting channels.

Preliminary results of this work were presented on the M2S-HTSC-IV conference in
Houston\cite{houston}. 
Subsequent to completion of this work we learn about a recent preprint by D. Orgad 
{\it et al.} (cond-mat/0005457) and by M.G. Zacher {\it et al.} (cond-mat/0005473),
where the relationship between stripe patterns and ARPES data is discussed.
We thank A.A. Aligia, S. Kivelson, Z-X. Shen and X.J. Zhou for useful discussions. 
Work at Los Alamos is
sponsored by the US DOE under contract W-7405-ENG-36.

\end{multicols}
\end{document}